\documentclass[a4,usenatbib,fleqn]{mn2e}
\usepackage{mathptmx}
\usepackage{amsmath,amssymb,amsbsy}
\usepackage[pdftex]{graphicx}
\usepackage[backref=none]{hyperref}
\usepackage[usenames,dvipsnames]{color}
\usepackage{bm}
\usepackage{txfonts}
\DeclareSymbolFont{cmletters}{OML}{cmm}{m}{it}
\DeclareMathSymbol{v}{\mathalpha}{cmletters}{"76}
\def\gcc{\hbox{\rm\hskip.35em  g cm}$^{-3}$}
\def\rads{\hbox{\rm\hskip.35em rad s}$^{-1}$}
\def\radss{\hbox{\rm\hskip.35em  rad s}$^{-2}$}
\def\cms{\hbox{\rm\hskip.35em  cm s}$^{-1}$}
\def\dcm{\hbox{\rm\hskip.35em  dyne s}$^{-1}$}

\def\ss{\hbox{\rm\hskip.35em s}$^{-2}$}
\def\sss{\hbox{\rm\hskip.35em s}$^{-3}$}
\def\fmc{\hbox{\rm\hskip.35em fm}$^{-3}$}
\def\cmss{\hbox{\rm\hskip.35em  cm$^{2}$ s}$^{-1}$}

\definecolor{MyDarkBlue}{rgb}{0,0.1,0.7}
\hypersetup{pdfborder={0 0 0},colorlinks,breaklinks=true,
  urlcolor={blue},citecolor={MyDarkBlue},linkcolor={magenta}}

\newcommand{\abs}[1]{{\left|#1\right|}}

\newcommand{\apj}{ApJ}
\newcommand{\apjl}{ApJ}

\newcommand{\mnras}{MNRAS}
\newcommand{\nat}{Nature}

\newcommand{\aap}{A{\&}A}
\newcommand{\araa}{ARA{\&}A}

\newcommand{\apss}{Ap{\&}SS}

\newcommand{\prc}{Phys. Rev. C}
\newcommand{\prd}{Phys. Rev. D}

\newcommand{\ptp}{Prog. Theoret. Phys. Suppl.}

\title[Vela glitch and NS interior dynamics]{The 2016 Vela glitch: a key to neutron star internal structure and dynamics}

\author[G\"{u}gercino\u{g}lu \& Alpar]{Erbil G\"{u}gercino\u{g}lu$^{1}$\thanks{Contact e-mail: \href{mailto:erbil.gugercinoglu@sabanciuniv.edu}{erbil.gugercinoglu@sabanciuniv.edu}} and M. Ali Alpar$^{1}$\thanks{Contact e-mail: \href{mailto:ali.alpar@sabanciuniv.edu}{ali.alpar@sabanciuniv.edu}}
\\
$^{1}$Faculty of Engineering and Natural Sciences, Sabanc{\i} University, Orhanl{\i}, Tuzla, 34956 Istanbul, Turkey}

\date{Accepted XXX. Received YYY; in original form ZZZ}

\pubyear{2020}

\begin{document}
\label{firstpage}
\pagerange{\pageref{firstpage}--\pageref{lastpage}}
\maketitle

\begin{abstract}
High resolution, pulse to pulse observation of the 2016 Vela glitch and its relaxation provided an opportunity to probe the neutron star internal structure and dynamics with unprecedented detail. We use the observations of this glitch to infer superfluid characteristics in the framework of the vortex creep model. The glitch rise time constraint of 12.6 seconds put stringent limits on the angular momentum exchange between the crustal superfluid and the observed crust. Together with the observed excess acceleration in the rotation rate as compared to the post-glitch equilibrium value this discriminates crustal superfluid-crust lattice and core superfluid-crustal normal matter coupling time-scales. An evident decrease in the crustal rotation rate immediately before the glitch is consistent with the formation of a new vortex trap zone that initiates the large scale vortex unpinning avalanche. Formation of vortex trap by a crust breaking quake induces short-lived magnetospheric changes. The long term post-glitch spin-down rate evolution reveals the moments of inertia and recoupling time-scales of the superfluid layers participating in the glitch and leads to an estimation of the time to the next glitch which agrees with the time interval between the 2016 and 2019 glitches. Our results are consistent with theoretical estimates of effective neutron and proton masses in the superfluid. We also constrain the vortex line-flux tube pinning energy per intersection as 2 MeV.
\end{abstract}





\section{Introduction} \label{sec:intro}

The Vela pulsar, with large glitches $\Delta\nu/\nu\sim10^{-6}$ at intervals of $\sim1000$ days, has been an 
emblematic source for establishing and testing glitch models thanks to dedicated monitoring 
programs  \citep{cordes88,buchner08,shannon16,palfreyman18}. 
Until recently glitch models took neutron stars as two component systems. One component in these models is the 
crustal superfluid, which is the agent of the glitch itself as well as the site of post-glitch recovery. 
The other component is the normal matter which consists of crustal normal matter, electrons thoughout the star and 
effectively includes the superfluid core, which is coupled to the normal matter on time-scales shorter than the 
observational bounds for the glitch rise time. Until the 2016 Vela glitch, 
best constraints for the glitch rise time were obtained for the 2000 and 2004 glitches, 
with upper limits of 40 s and 30 s, respectively \citep{dodson02,dodson07}. High resolution, pulse to pulse 
observations of the 2016 Vela glitch \citep{palfreyman18} have brought this upper limit to much lower values. 
\citet{ashton19}  used these data and employed Bayesian techniques to bring a 12.6 seconds upper limit to the glitch 
rise time at 90\% confidence level. Their analysis also resolved the peak of the initial glitch increase in the observed
 crustal angular velocity which decays to a lower level with a $\lesssim 1$ minute exponential relaxation time-scale immediately after the glitch. This promptly 
decaying component contains more than half of the total glitch amplitude. Resolved promptly decaying components 
were barely evident for the 2000 and 2004 glitches. 

These results point to a three component neutron star model we consider: angular momentum is first transferred from 
crustal superluid to the crustal normal matter in less than 12.6 s and thereafter shared with the core superfluid 
within a minute. Other three component models for glitch dynamics \citep{graber18, pizzochero19, sourie20} 
employ different assignments of the agents of angular momentum transfer to components of the neutron star 
and/or different couplings between them compared to the specific model we present here.

\citet{ashton19} also reported definitive evidence for an apparent decrement in the crustal rotation 
rate right before the glitch, a behaviour never resolved previously from any glitches. 
In this paper we evaluate these observations in terms of the vortex creep model and explore a glitch scenario 
accounting for them. We propose that the crustal rotational velocity decrease prior to the 2016 glitch 
marks the formation of a new vortex trap zone, and that the peak glitch amplitude and its prompt relaxation are 
signatures of glitch rise due to fast coupling of the crustal superfluid first to the crustal normal matter, which is followed by the 
gradual coupling of crustal normal matter to the core superfluid. 
In \S\ref{sec:observations}  we summarize the main observational features of the 2016 Vela glitch. 
In \S\ref{sec:theory} we propose a scenario within the vortex creep model and then obtain constraints on the model parameters. 
We discuss our results in \S\ref{sec:dandc}.

\section{Observational Features of the 2016 Vela Glitch} \label{sec:observations}

The study of the detailed analysis of the 2016 Vela glitch is based on single pulse to pulse observations conducted at 
Mount Pleasant radio telescope by \citet{palfreyman18}. This glitch occurred on 12 December 2016. 
The fractional changes in the pulsar frequency and spin-down rate are $\Delta\nu/\nu=1.431(2)\times10^{-6}$ 
and $\Delta\dot\nu/\dot\nu=73.354\times10^{-3}$, respectively \citep{xu19}. \citet{palfreyman18} detected 
pulse morphology and polarization level variations starting 20 rotations before the glitch and extending through the 
first 4.4 seconds after the glitch. They interpreted this observation as a transient change in the magnetospheric state associated 
with the glitch. \citet{ashton19} reanalysed observational data presented by \citet{palfreyman18} with Bayesian 
techniques.   \citet{xu19} reported on the longer term post-glitch $\Delta\nu(t)$ and $\Delta\dot\nu(t)$ recoveries 
based on analysis of timing data from the Kunming 40-m radio telescope. \citet{basu20} also 
reported on the $\Delta\nu(t)$ recovery of this glitch with timing data from the Ooty radio telescope.
These observations and analysis led to the following conclusions:

\begin{enumerate}
\item Prior to the glitch there was a decrease in the rotation rate by $\Delta\nu_{-}=5.40^{+3.39}_{-2.05}$ $\mu$Hz 
for $\sim100$ seconds, with a significance that
is difficult to establish \citep{ashton19}. This magnitude is comparable to the glitch size itself. 
\item There was a temporary change in pulse shape and one missed pulse at the time of the glitch \citep{palfreyman18}.
\item The tightest limit so far was obtained for the glitch rise time $t_{\rm rise} < 12.6$ s \citep{ashton19}.
\item The peak initial spin-up in the glitch, $\Delta\nu_{\rm d+}=17.77^{+13.68}_{-7.99}$ $\mu$Hz promptly relaxed with 
an exponential decay time $\tau_{\rm d+}=53.96^{+24.02}_{-14.82}$ seconds \citep{ashton19}.
\item After the prompt decay of the initial peak, the remaining frequency step of $\Delta\nu=16.01(5)$ $\mu$Hz relaxed 
with two short time-scale exponential decay terms with time constants 1 and 6 days, 
and a long term healing of $\Delta\dot\nu(t)$ with a constant $\Delta\ddot\nu$ ensued \citep{xu19}. 
\end{enumerate}

The Vela pulsar glitched once again on 1 February 2019 \citep{sarkissian19,gancio20}.

\section{Vortices and Angular Momentum Exchange} \label{sec:theory}

Elements of the vortex unpinning and vortex creep model for glitches are, in time order: 
\begin{enumerate}
\item Possible vortex trap formation and quake triggered  events lead to the vortex unpinning avalanche. 
\item Crust breaking as trigger may have magnetospheric signatures like pulse shape and emission behaviour changes.
\item Glitches themselves are vortex unpinning events which first transfer angular momentum from crustal superfluid 
to the normal matter nuclei and electrons in the crust. 
\item The crustal superfluid plus normal matter in the crust then couple via electrons to the core superfluid 
on still very short time-scales.
\item Once the core superfluid is coupled to the normal matter in the crust, the core superfluid + normal matter system 
behaves as an effective crust which contains most of the moment of inertia of the star. 
This effective “crust” relaxes back with crustal superfluid as the continuous vortex creep process builds up again towards 
the steady-state pre-glitch conditions.
\end{enumerate}

All except (v) were not observed in the Vela pulsar before. (i) trap formation and triggering quake were 
surmised in the Crab pulsar \citep{erbil19} and (ii) glitch induced magnetospheric changes observed for
PSR J1119$-$6127 \citep{akbal15}.  So far for the Vela and all other pulsars only (v) the post-glitch recovery was 
fitted with creep response models.

\subsection{Formation of vortex traps leading to crustal rotation rate decrease and a quake triggering the glitch} \label{sec:trap}

The observed decrement in the crustal rotation rate prior to the glitch is one of the striking properties of the 2016 Vela 
glitch. Such a behaviour had never been seen before from any pulsar glitch. The 2000 and 2004 Vela pulsar glitches  
\citep{dodson02, dodson07} had the previous smallest 
uncertainties in the glitch occurrence time. For the former glitch high cadence arrival time data were not available prior 
to the glitch, while for the latter the immediate pre-glitch data were noisy.   

Glitches involving superfluid vortex unpinning may be triggered by crustquakes. We propose that the slow-down prior to 
the 2016 glitch is a signal of the formation of a new vortex trap in association with crust breaking, which then provided 
the site where the glitch was triggered. The motion of broken crustal plates would lead to extra vortex pinning, creation of vortex 
free regions and induced motion of clusters of pinned vortices. This idea was first proposed by \citet{alpar96} in order 
to account for the persistent shifts from the Crab glitches \citep{lyne93}, i.e. the glitch associated 
permanent increases in the observed spin-down rate which do not recover subsequently. 
Further support for the glitch trigger involving both crust breaking and its 
induced effects on the configuration of pinned vortices came from the realization that neither pure crustquake models 
nor pure vortex unpinning and creep recovery models could explain the intervals between the Crab pulsar
glitches \citep{alpar96}. This idea was further applied to the Crab pulsar's largest glitch by \citet{erbil19}. 

Vortex density in the pinned superfluid is unlikely to be uniform. 
Depending on the distribution of pinning centres of various strengths, and on the history of the pinned superfluid, 
vortex traps which are high vortex density regions surrounded by vortex depletion regions 
will be formed, as first suggested by \citet{cheng88}. A vortex trap is formed when the local lag between the crustal superfluid and the crust angular velocities increases from the steady-state lag $\omega_{\infty}$ to a value above the critical lag $\omega_{\rm cr}$ for unpinning or re-pinning, thereby allowing for extra 
vortex pinning in the trap. The extra (repinned) vortex density leads to the formation of a vortex free region as part of the trap: 
Due to the extra pinned vortices the local superfluid velocity in the surrounding regions becomes so large that 
the lag is sustained at a value above the critical value for unpinning and no pinning sites are effective in the 
surrounding region. Therefore, the vortex creep process 
is not sustained in the newly and irreversibly formed vortex free part of the trap, leading to the observed slow-down 
of the crust before the glitch. 

\begin{table}
\caption{Pinning parameters and increase in the lag needed for unpinning in various layers of the crustal superfluid. Entries in the first three columns are taken from \citet{seveso16} ($\beta=3$ model). The last column is calculated from Eq. (\ref{steadyunpin}) using a temperature $ kT=6.5$ keV and $v_{\rm 0}$ values from \citet{erbil16}.}
\label{pinpar}
\begin{center}{
\begin{tabular}{cccc}
\hline\hline\\
\multicolumn{1}{c}{$\rho$} & \multicolumn{1}{c}{$E_{\rm p}$} & \multicolumn{1}{c}{$f_{\rm p}$} & \multicolumn{1}{c}{$\omega_{\rm cr}-\omega_{\infty}$} \\
\multicolumn{1}{c}{($10^{13}\mbox{\gcc}$)} & \multicolumn{1}{c}{(MeV)} & \multicolumn{1}{c}{($10^{14}\mbox{\dcm}$)} & \multicolumn{1}{c}{($\mbox{\rads}$)} \\ 
\hline\\
0.15 & 0.21 & 3.2 & 0.101 \\\\
0.96 & 0.29 & 3.1 & 0.012 \\\\
3.4 & 2.74 & 85.5 & 0.011 \\\\
7.8 & 0.72 & 18.4 & 0.004 \\\\
13 & 0.02 & 0.6 & 0.003 \\\\
\hline\\
\label{vortextrap}
\end{tabular}}
\end{center}
\end{table}

For the change $\delta\Omega_{\rm s, trap}$ in superfluid rotation rate in the newly formed 
vortex trap-vortex free regions and in turn decline of the angular momentum transfer rate to the observed crust we use the estimate \citep{erbil16}
\begin{equation}
\delta\Omega_{\rm s, trap}\cong\omega_{\rm cr}-\omega_{\infty}=\frac{f_{\rm p}}{\rho\kappa R}\frac{kT}{E_{\rm p}}\ln\left(\frac{2\Omega_{\rm s}v_{\rm 0}}{\abs{\dot\Omega}R}\right),
\label{steadyunpin}
\end{equation}
where $T$ is the temperature of the inner crust, $\kappa \cong 2\times 10^{-3}$\mbox{\cmss} is the quantum of vorticity, $\Omega_{\rm s}$ is the superfluid 
angular velocity ($\approx \Omega_{\rm c}$, the rotation rate of the neutron star since the lag $\omega$ is small), 
$f_{\rm p}$ is the pinning force per unit length of a vortex line, $E_{\rm p}$ is the pinning energy, $\rho$ is the matter density, 
$R$ is the vortex line's distance from the rotation axis, $\abs{\dot\Omega}$ is the magnitude of the spin-down rate, and $v_{\rm 0}$ is the microscopic vortex velocity around nuclei. 
The permanent establishment of a vortex free region means that the corresponding region no longer transfers superfluid angular momentum 
to the crust because it no longer sustains vortices or vortex creep. The crust is then spinning down at a slightly 
higher rate leading to a decrease $\Delta\nu_{-}$ in the observed spin frequency of the crust
\begin{equation}
\Delta\nu_{-} =\frac{\Delta\Omega_{\rm c-}}{2\pi} =\frac{1}{2\pi}\frac{I_{\rm trap}}{I}\left(\omega_{\rm cr}-\omega_{\infty}\right),
\label{aveltrap}
\end{equation}
due to the formation of a vortex trap with fractional moment of inertia $I_{\rm trap}/I$. 
From Eqs. (\ref{aveltrap}) and (\ref{steadyunpin}) with $ kT=6.5$ keV from observations of the Vela \citep{vigano13} and 
entries in Table \ref{vortextrap} we obtain
\begin{equation}
\frac{I_{\rm trap}}{I}=2.08\times10^{-4}-1.98\times10^{-2},
\label{newtrap}
\end{equation}
for the observed value $\Delta\nu_{-}=5.40^{+3.39}_{-2.05}$ $\mu$Hz \citep{ashton19}.

The establishment of a new vortex trap is a redistribution of vortices, by typical microscopic motions over the 
distance to nearby vortices at the average vortex spacing $\ell_{\rm v}=(2\Omega_{\rm s}/\kappa)^{-1/2}$. 
The time-scale for the trap formation and the accompanying persistent increase in the spin-down rate and 
decrease in the rotation rate of the crust is then the time taken by a vortex to move one inter-vortex 
spacing $\ell_{\rm v}$ at the typical microscopic 
speeds $v_{0}$ and is given by \citep{alpar89}
 \begin{equation}
t_{\rm tr}=\frac{\ell_{\rm v}}{v_{0}}\exp\left(\frac{E_{\rm p}}{kT}\right).
\label{vtransit}
\end{equation}
By taking $v_{\rm 0}\approx 10^{7}$\mbox{\cms} \citep{erbil16}, $ kT=6.5$ keV \citep{vigano13} 
and for $E_{\rm p}=0.17$ MeV pertaining to densities $\rho\cong 10^{14}$\gcc , ~Eq. (\ref{vtransit}) gives $t_{\rm tr}\cong 100$ s. 
This agrees with the observed time-scale of the immediate pre-glitch slow-down epoch \citep{ashton19}. Having determined the location of the trap region with $E_{\rm p}=0.17$ MeV corresponding to densities $\rho\cong10^{14}$\gcc~the fractional moment of inertia of the newly formed trap constraint (\ref{newtrap}) now becomes $I_{\rm trap}/I=8.58^{+5.38}_{-3.26}\times10^{-3}$ for $\Delta\nu_{-}=5.40^{+3.39}_{-2.05}$ $\mu$Hz.

Since Eq. (\ref{vtransit}) depends exponentially on the pinning energy $E_{\rm p}$ of the crustal region where 
the vortex avalanche starts, $t_{\rm tr}$ can be unobservably short for smaller $E_{\rm p}$. 
The pinning energy $E_{\rm p}$ decreases 
with increasing density at the higher densities in the inner crust \citep{alpar77,seveso16}. 
Glitches originating in the inner crust superfluid at layers denser and deeper than the 
location of the vortex trap triggering the 2016 glitch would have durations of the spin-down event immediately 
preceding the glitch that are (exponentially) shorter than $\sim 100$ s, thus becoming hard to detect such an event. Such glitches originating from deeper
layers would also be larger glitches as the unpinned vortices would be moving through 
more crustal superfluid layers and therefore cause more angular momentum transfer. 
This is consistent with the fact that the 2016 glitch, the first case of immediate pre-glitch data resolved at this level \citep{ashton19}, is one of the smaller glitches, below the median for Vela glitches \citep{xu19}. The Vela pulsar's larger glitches are likely to be originated from the inner crust at densities comparable to and larger 
than the density range $\rho\simeq 10^{14}$\mbox{\gcc}. If our interpretation is correct, then for the largest 
glitches of the Vela pulsar wherein vortex unpinning avalanche starts deeper in the inner crust and therefore involves larger 
non-linear creep superfluid moments of inertia $I_{\rm A}$ in the regions outward of the trigger, the pre-glitch 
slow-down episode lasts for on still shorter time-scales than is the case for the medium sized 2016 glitch which 
we estimate to have originated in an intermediate layer of the inner crust. Since the non-linear creep superfluid 
moments of inertia inferred for the 2000 and 2004 Vela glitches observed by \citet{dodson02,dodson07} 
[see Table 1 in \citet{akbal17}] are slightly larger than those inferred for the 2016 glitch (see Table \ref{fitmodel} below), 
the trigger region for those glitches should be deeper in the inner crust and hence characterized by time-scales 
much shorter than that of the slow-down event preceding the 2016 glitch, and therefore we expect the slow down phase right before these glitches observed by \citet{dodson02,dodson07} to be only barely apparent if at all. Glitches comparable to or smaller than the 2016 glitch in terms of fractional change in pulsar frequency 
will display pre-glitch spin-down episodes of 100s or longer duration if resolved.

\subsection{Effects of the glitch triggering quake on the magnetosphere}

The 2016 glitch was accompanied by short-lived changes in the Vela pulsar's electromagnetic signature. 
There were a pulse shape change (broadening) and notably a null state starting 3.3 s before the glitch with a missed pulse 
within 0.2 s, and a loss of linear polarization in the next two pulses extending less than 4.4 s at the 
time of the glitch \citep{palfreyman18}.

Changes in the pulsar's signature associated with the glitches are very rarely observed, so far only from 
the 2007 glitch of PSR J1119$-$6127 [except for the magnetar glitches \citep{kaspi17} and for minority of pulsars there are some circumstantial evidences, see for instance \citet{yuan19}]. If the broken crustal plate triggering 
the glitch happens to extend to the neutron star surface, the glitch will influence the magnetosphere through its coupling to 
the magnetic field lines anchored in the conducting crust, leading to changes in the electromagnetic signature of the pulsar 
and the dipole spin-down  torque. This idea was first applied by \citet{akbal15} to the 2007 peculiar glitch 
of PSR J1119-6127 \citep{weltevrede11, antonopoulou15}, which exhibited a clear change in the pulsar signature and 
a possible change in the dipole spin-down torque. The 2016  glitch is the first glitch observed from the Vela pulsar 
with a glitch precursor decrease in frequency as well as a change in the pulse morphology. 
\citet{bransgrove20} have pursued the idea of a crust breaking trigger for the 2016 Vela glitch and 
presented its consequences for the magnetospheric changes. In PSR J1119$-$6127 the 2007 glitch also 
led to magnetospheric activity change in the form of transient intermittency and RRAT
behaviour for about three months after which the pulsar resumed
to pre-glitch signature. \citet{akbal15} interpreted peculiarities associated with the 2007 glitch of PSR J1119$-$6127 as a 
result of crust breaking extending to the surface in the polar cap which brings about magnetospheric activity changes 
by twisting of the magnetic field lines on the scale of the broken plate motion, 10 - 100 m. This large length scale disturbance 
was associated by \citet{akbal15} with the 3 month duration of 
the glitch induced transient changes (intermittency and RRAT-like behaviours). 
\citet{bransgrove20} have proposed that the $\sim $ 4 seconds
short time-scale transient magnetospheric activity change coincident with the 2016 glitch may arise from 
ephemeral electron/positron discharge due to bouncing seismic waves and energy pumped into the magnetosphere 
at correspondingly high frequencies on short time and length scales
by a quake occurring deep in the crust. 

\subsection{Angular momentum transfer from superfluid to normal matter in the crust and the limit on the glitch rise time}

Glitches are thought to be arising from rapid angular momentum transfer from crustal superfluid to the crust mediated 
by vortex lines. One such mechanism is kelvon wave excitation on vortex lines and their coupling with the 
lattice phonons \citep{epstein92,jones92}. As shown by \citet{graber18} this mechanism leads to very effective 
coupling of the superfluid to the crust, consistent with the 12.6 seconds upper limit obtained by \citet{ashton19}. 
The crustal superfluid-crustal normal matter coupling time-scale can be expressed in terms of a drag coefficient $\Re$ by
\begin{equation}
\tau_{\rm mf}=\left(\frac{1+\Re^{2}}{\Re}\right)\frac{I_{\rm cs}/I_{\rm cn}}{2\Omega},   
\label{kt}
\end{equation}
where $I_{\rm cs}$ and $I_{\rm cn}$ are the moments of inertia of the crustal superfluid and the normal matter in the crust, 
respectively. \citet{delsate16} have obtained the ratio $I_{\rm cs}/I_{\rm cn}\cong8.35$. 
For kelvon wave coupling, the drag coefficient $\Re$ can be expressed in terms of the crustal parameters as \citep{graber18,erbil16}
\begin{equation}
\Re=1.14\left(\frac{v_{0}}{10^{7}\mbox{\cms}}\right)^{-1/2}\left(\frac{a}{63\mbox{fm}}\right)^{-3}\left(\frac{R_{\rm N}}{7\mbox{fm}}\right)^{5/2}. 
\label{dragk}
\end{equation}
Here $v_{0}$ is the microscopic vortex line velocity around nuclei, $R_{\rm N}$ is the nuclear radius and $a$ is the 
lattice constant. The range  of $v_{0}$  values was most recently obtained by \citet{erbil16}. For instance, 
at baryon density $n_{\rm B}=7.89\times10^{-2}$\fmc, the corresponding parameters 
are $a=29.2$ fm, $R_{\rm N}=7.2$ fm, $v_{0}=4.58\times10^{6}$\cms ~yielding $\tau_{\rm mf-kelvon}\approx 1.1$ s; 
while for the lower density $n_{\rm B}=3.73\times10^{-3}$\fmc, $a=81.2$ fm, $R_{\rm N}=6.6$ fm, 
and $v_{0}=6.4\times10^{5}$\mbox{\cms} Eqs. (\ref{kt}) and (\ref{dragk}) give $\tau_{\rm mf-kelvon}\approx 0.14$ s. 
Thus, throughout the neutron star crust kelvon waves bring about very effective coupling on 
time-scales of order $\tau_{\rm mf-kelvon}\sim 0.1-1$ s. A general relativistic treatment is likely to lead 
to even tighter coupling \citep{sourie17, gavassino20}. This very fast coupling implies that we will not be able to discriminate the coupling 
between the crustal superfluid and the normal matter crust components through post-glitch timing observations. This is because
the construction of pulse templates used in time of arrival analysis involves binning of many individual cycles. 
Timing analysis for the Vela pulsar is possible on time-scales longer than at least a few seconds  \citep{palfreyman18}. 
Short $\tau_{\rm mf-kelvon}$ also sets the scale for the detection of transient gravitational wave emission 
associated with the glitches if this becomes feasible with future instruments \citep{melatos15,keitel19}.

\citet{pizzochero19} fit phase residual data right after the glitch from \citet{palfreyman18} by their three component model and obtained glitch rise time as $0.20\pm 0.14$ s which is in line with our estimate $\tau_{\rm mf-kelvon}\sim 0.1-1$ s.  

\citet{celora20} considered the effect of  non-linear relative velocity dependence of the mutual friction force between the vortex lines and the normal matter on the glitch rise time, finding that inverse power law dependence on the relative velocity, as in our case with Eq.(\ref{dragk}), leads to faster glitch rise.

\citet{sourie20} considered 12.6 seconds upper limit as the glitch rise time with the interpretation that angular momentum transfer occurs as a consequence of unpinning of the vortex lines from magnetic flux tubes in the outer core.

\subsection{Prompt relaxation of the peak spin-up due to crust-core coupling}

From the earliest on Vela pulsar glitch in 1969, observations showed that post-glitch relaxation involves only at most a few 
percent of the moment of inertia of the star on time-scales longer than hours. The implication was that the bulk of 
the star's moment of inertia, which lies in the core superfluid was already coupled to the crust on time-scales less than a minute. 
This is explained by spontaneous magnetization of vortex lines in the core neutron superfluid by dragging superconducting 
proton currents, leading to very effective electron scattering off spontaneous magnetized vortex lines \citep{ALS84}.  
Thanks to the high resolution, pulse to pulse observations of the 2016 Vela glitch \citep{palfreyman18} it now becomes possible
to distinguish items (iii) and (iv) in Section \ref{sec:observations}. The former corresponds to the crustal superfluid-normal matter crust coupling which must have taken place within $t_{\rm rise}= 12.6~\rm{s}$. We associate the latter, 
the prompt  $\tau_{\rm d+}\cong 54~\rm{s}$ decay following the initial peak of the glitch, with crust-core coupling taking place by sharing the initial angular 
momentum imparted to the crustal normal matter with the core superfluid on the time-scale $\tau_{\rm core}$  of electron scattering from
spontaneously magnetized vortex lines \citep{ALS84}. 

Thus, the glitch is observed on the shortest time-scales in a sequence of angular momentum transfers 
between the three components: the crustal superfluid that presents a vortex unpinning event, the crust normal matter 
as the first recipient of the angular momentum transfer and the core superfluid which subsequently takes part in sharing of 
the angular momentum. The crustal rotation rate increase $\Delta\Omega_{\rm c}(0)$ at the time of the glitch is determined from the angular momentum exchange between the crustal superfluid (with moment of inertia $I_{\rm cs}$) and the crustal normal matter (with moment of inertia $I_{\rm cn}$) first, and is given by
\begin{equation}
\frac{\Delta\Omega_{\rm c}(0)}{\Omega_{\rm c}}=\frac{I_{\rm cs}}{I_{\rm cn}}\frac{\delta\Omega_{\rm s}}{\Omega_{\rm c}},
\label{gexcess}
\end{equation}
with $\delta\Omega_{\rm s}$ being the change in the superfluid angular velocity due to unpinning of vortex lines. 
After the sharing of the glitch spin-up with the core superfluid on the relaxation time-scale $\tau_{\rm core}$ 
given by Eq. (\ref{ALStime}) below, the
immediate post-glitch magnitude of the ``effective" crust comprised of the crust and the core superfluid 
is \citep{erbil14}
\begin{equation}
\frac{\Delta\Omega_{\rm c}}{\Omega_{\rm c}}=\frac{fI_{\rm cs}}{I-I_{\rm cs}-I_{\rm tor}}\frac{\delta\Omega_{\rm s}}{\Omega_{\rm c}},
\label{gequilibrium}
\end{equation}
where the factor $f\lesssim1$ designates the fraction of the crustal superfluid that participated in the glitch event. 

At times between glitches, the crustal superfluid where vortex lines can be pinned to 
nuclei [as well as the superfluid in the outer core where vortex lines can pin to a toroidal arrangement of quantized flux lines \citep{erbil14}] 
support continuous angular momentum transfer by vortex creep.  After the motion of unpinned vortices during the glitch,
the conditions driving creep are deeply offset leading to temporary decoupling of the creep regions in the crust (with moment of inertia $I_{\rm cs}$) and the outer core (with moment of inertia $I_{\rm tor}$) from the whole neutron star (with moment of inertia $I$) on longer time-scales.

\begin{figure}
\centering
\vspace{0.1cm}
\includegraphics[width=1.0\linewidth]{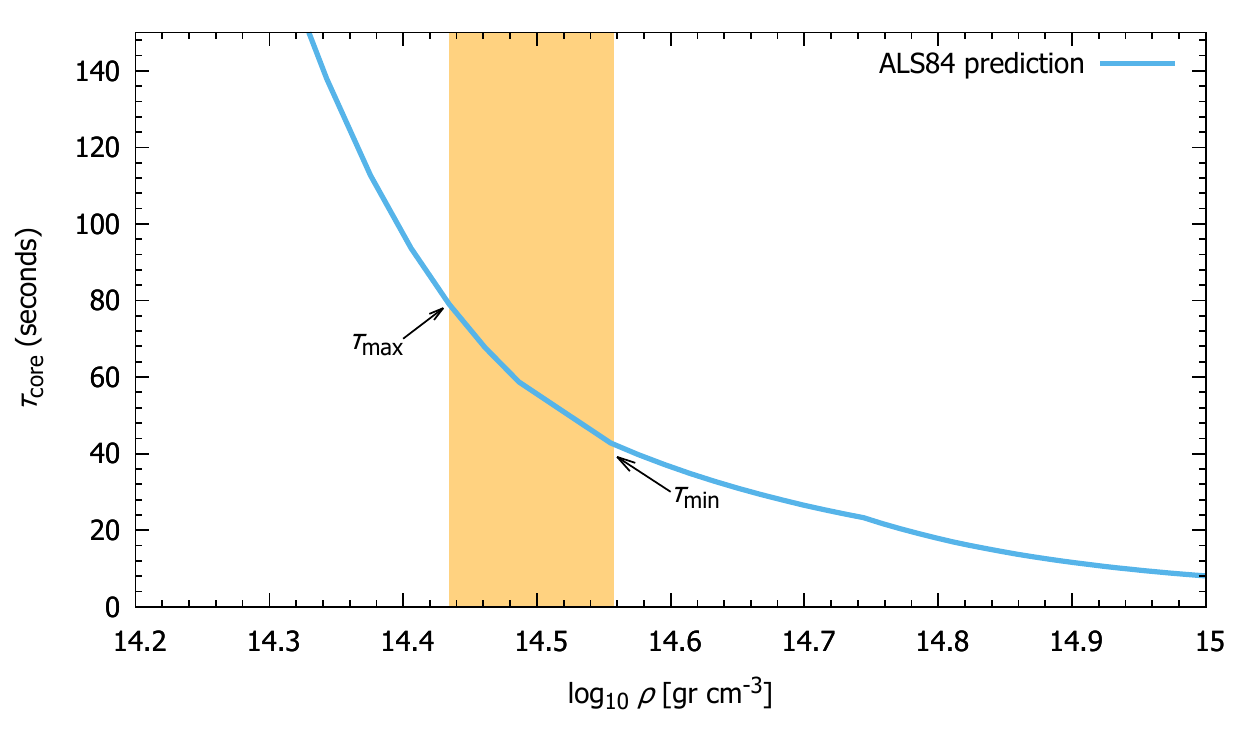}
\caption{Crust-core coupling time predicted by Eq. (\ref{ALStime}) versus matter density. 
The shaded region bounded with arrows corresponds to the interval of 
values $\tau_{\rm d+}=53.96^{+24.02}_{-14.82}$ s obtained by \citet{ashton19} from timing data.}
\label{ALScoupling}
\end{figure}

The crust-core coupling time-scale is given by \citep{sidery09, ALS84}
\begin{equation}
\tau_{\rm core}=6.2x^{-1/6}_{\rm p}\left(\frac{\rho}{10^{14}\mbox{\gcc}}\right)^{-1/6}\left(\frac{\delta m^{*}_{\rm p}}{m_{\rm p}}\right)^{-2}\left(\frac{m^{*}_{\rm p}}{m_{\rm p}}\right)^{1/2}P,
\label{ALStime}
\end{equation}
where $x_{\rm p}$ is the proton fraction, $\rho$ is the mass density, $m^{*}_{\rm p}(m_{\rm p})$ 
is the effective (bare) proton mass, $\delta m^{*}_{\rm p}=m_{\rm p}-m^{*}_{\rm p}$ 
and $P$ is the rotational period in seconds. As angular momentum transfer between the core 
and the observed crust is weighed by the moment of inertia $dI \propto \rho(r) r^4 \sin^2(\theta)$ 
in spherical coordinates $(r,\;\theta\;\phi)$ with respect to the rotation axis, the effective 
coupling time will be dominated by the outer core layers with $\rho \sim 2\times10^{14}$\mbox{\gcc}. 
Comparing Eq.(\ref{ALStime}) with the relaxation time fitted to the observations 
yields  $m^{*}_{\rm p}/m_{\rm p} \sim 0.7$ at $\rho \gtrsim 2 \times 10^{14}$\mbox{\gcc} in agreement with 
theoretical calculations in this density range. Figure \ref{ALScoupling} shows the run 
of $\tau_{\rm core}(\rho)$ using the $m^{*}_{\rm p}/m_{\rm p}$ results of \citet{chamel08} and 
employing the A18 + $\delta v$ + UIX* equation of state of \citet{akmal98}. 
The prompt relaxation time-scales dicussed in item (iv) of Section \ref{sec:observations} are indicated with arrows. 
Relaxation time $\tau_{\rm d+}=53.96^{+24.02}_{-14.82}$ s corresponds to the core superfluid with 
densities $\lesssim4\times 10^{14}$\gcc. As can be seen from Figure \ref{ALScoupling}, for core regions with 
densities above $7.4\times 10^{14}$\gcc~the  crust-core coupling times are shorter than the glitch rise time constrained 
to $<12.6$ seconds, and thus were already coupled to the normal matter crust when the glitch was resolved. 
 The amplitude of the prompt decay component of the  
glitch can be expressed in line with Eqs. (\ref{gexcess}) and (\ref{gequilibrium}) as
\begin{equation}
\frac{\Delta\nu_{\rm d+}}{\Delta\nu}=\frac{I'_{\rm core}}{I_{\rm c}-I'_{\rm core}},
\label{deltaIcore}
\end{equation} 
where $I_{\rm c}$ is the moment of inertia of the core superfluid already coupled to the observed crust on a time-scale 
shorter than the observational upper limit of 12.6 seconds, while $I'_{\rm core}$ denotes 
the moment of inertia of the core regions which couple to the crust through the observed prompt decay. 
With $\Delta\nu_{\rm d+}=17.77^{+13.68}_{-7.99}$ $\mu$Hz and $\Delta\nu=16.01(5)$ $\mu$Hz 
we obtain from Eq. (\ref{deltaIcore}) the constraint $I'_{\rm core}\cong(0.38-0.66)I_{\rm c}$. 
This rather large fraction corresponding to the shaded region in Figure \ref{ALScoupling} is consistent 
with the fact that $I\propto \rho R^{5}$.

 \citet{pizzochero19} interpreted excess rotational increase (which is greater than the mean post-glitch equilibrium value) immediately following the glitch as angular momentum exchange between three components, namely crustal superfluid, core superfluid and non-superfluid core matter. 

\citet{sourie20} are able to account for both initial peak in crustal angular velocity and its subsequent quick recovery within their model assumptions by adjusting number of vortex line-flux tube pinning sites as a free parameter.

\subsection{The long term recovery and fit with the vortex creep model}

The post-glitch relaxation on time-scales of days or longer are modelled on the basis of the response of vortex creep in the pinned 
superfluid to the glitch. We begin this section by a summary of the vortex creep. 

The standard model for pulsar glitches \citep{anderson75} invokes pinning of the superfluid vortex lines to the 
crustal lattice in order to develop an angular velocity difference between the normal matter crust and crustal superfluid, 
thereby sustaining a reservoir of angular momentum. Glitches occur whenever and wherever the angular velocity 
lag $\omega=\Omega_{\rm s}-\Omega_{\rm c}$ exceeds the critical threshold $\omega_{\rm cr}$ for unpinning of 
vortex lines. Collective unpinning of vortices initiates an avalanche which bring about a glitch as confirmed by numerical 
simulations \citep{warszawski11,melatos19}. In between glitches these vortex lines cannot remain absolutely pinned. 
At finite temperature, vortex lines have probabilities proportional to Boltzmann 
factors to overcome the pinning potential barriers and jump between adjacent pinning sites, with a bias for radially outward 
motion, as dictated by spin-down of the pulsar under the external braking torque. 
The superfluid manages to spin down in the presence of pinning 
as a result of the flow of the vortex lines thermally activated against pinning energy barriers $E_{\rm p}$. This process is called ``vortex creep" \citep{alpar84}. 

The dynamical coupling between the superfluid and the crust provided by vortex creep is an exponential 
function of the lag $\omega$, which acts as the driving force for the vortex current,  analogous to the voltage in electric circuits.  
Glitches due to sudden vortex unpinning (discharge) \citep{anderson75} are analogous to capacitive discharges 
superposed on and interacting with the continuous process of vortex creep. 
In some parts of the pinned superfluid the response of vortex creep is linear in the 
glitch induced changes, leading to exponential relaxation \citep{alpar89,erbil14}. 

Most parts of the pinned superfluid sustain non-linear creep with a response that is highly non-linear in the glitch induced changes. 
In the conventional notation of papers on glitches and creep \citep{alpar84,alpar89,alpar96}, the moment of inertia of the non-linear creep regions
that are affected by the glitch is denoted by $I_{\rm A}$. The notation $I_{\rm B}$ is used for the moment of 
inertia of the vortex free ``capacitive" regions surrounding the vortex traps. Like the space between capacitor plates, 
such vortex free regions do not sustain any vortices and therefore do not contribute to the vortex creep ``current". 
As discharged vortices move through these regions in the glitch, 
they contribute to the angular momentum transfer to the crust, observed as the glitch, in proportion to 
the moment of inertia $I_{\rm B}$. 

A glitch involves unpinning of a very large number of 
vortices whose sudden motion through the superfluid decreases the superfluid angular velocity 
by $\delta\Omega_{\rm s}$ and transfers angular momentum to the crust normal matter which spins up 
by $\Delta\Omega_{\rm c}$. The effect of the glitch on the steady-state angular velocity lag $\omega_{\infty}$, which is the driving force of the vortex motion (analogous to voltage
in an electronic circuit), and thereby on the creep current is analogous to what happens to the behaviour of  charge carriers (current) in electric circuits when the voltage drops: In the aftermath of the glitch the angular velocity lag has suddenly decreased by $\delta\omega=\delta\Omega_{\rm s}+\Delta\Omega_{\rm c}$. Then, the creep process itself, which allows the superfluid to spin-down and transfer angular momentum to the crust, weakens; indeed non-linear creep, which depends very sensitively on the
angular velocity lag, can temporarily stop in the corresponding crustal superfluid region.
With the corresponding superfluid region decoupled from the crust, the external torque 
is acting on less moment of inertia so that the spin-down rate increases. This behaviour persists until 
steady-state creep conditions are reestablished after a waiting time $t_{0}= \delta\Omega_{\rm s}/\abs{\dot\Omega}$.

In the Vela pulsar, after the exponential recoveries are over, the observable quantities associated with 
the glitches are related to the vortex creep model parameters by
the following basic equations \citep{alpar06}:
\begin{equation}
I_{\rm c}\Delta \Omega_{\rm c} = (I_{\rm A}/2 + I_{\rm B}) \delta \Omega_{\rm s},
\label{glitchomega}
\end{equation}
\begin{equation}
\frac{\Delta \dot \Omega _{\rm c}(t)}{\dot \Omega _{\rm c}}=\frac{I_{\rm A}}{I_{\rm c}}\left(1-\frac{t}{t_{0}}\right),
\label{glitchdotomega}
\end{equation}
\begin{equation}
\Delta\ddot{\Omega} _{\rm c} = \frac{I_{\rm A}}{I_{\rm c}} \frac{{\dot\Omega}^2}{\delta\Omega_{\rm s}}.
\label{ddotomega}
\end{equation}
The model parameters $I_{\rm A}$, $I_{\rm B}$, and $\delta\Omega_{\rm s}$ can be obtained from these equations on using the 
observed glitch parameters (in the long
term after the early exponential recoveries are over) without making a detailed fit to the data.

\citet{xu19} fit the long-term post-glitch data for 416 days starting 2 days after the glitch with the following function:
\begin{align}
\Delta\nu(t) =& \Delta\nu_{\rm d1}e^{-t/\tau_{\rm d1}}+\Delta\nu_{\rm d2}e^{-t/\tau_{\rm d2}}+\Delta\nu_{\rm p}+\Delta\dot\nu_{\rm p}t+\Delta\ddot\nu_{\rm p}\frac{t^{2}}{2}.
\label{modeleq}
\end{align}
Here $\Delta\nu =\Delta\nu_{\rm d1}+\Delta\nu_{\rm d2}+\Delta\nu_{\rm p}$ is the component of the glitch, excluding the 
prompt relaxation, as discerned and extrapolated from the data starting at  $t$ = 2 days after the glitch. 
Results of their data fit are included in Table \ref{fitmodel}. Their fit contains two exponential decay terms with 
time constants 1 and 6 days, respectively, and a decrease in $\abs{\dot\nu}$  
with a constant positive $\ddot\nu$. Spin-down rate evolution with constant, positive, 
large $\ddot\nu$ is a standard feature of the inter-glitch behaviour of the Vela pulsar \citep{alpar84,akbal17}. 
It is also common in the inter-glitch timing behaviour of other pulsars with large glitches \citep{johnston99,alpar06,yu13,dang20}.
Using the long term post-glitch timing fit parameters of \citet{xu19}, we obtain 
from Eqs. (\ref{glitchomega}), (\ref{glitchdotomega}), and (\ref{ddotomega}) the model parameter values
\begin{equation}
\frac{I_{\rm A}}{I_{\rm c}}=(6.42\pm 1.93)\times10^{-3},
\end{equation}
\begin{equation}
\frac{I_{\rm B}}{I_{\rm c}}=(1.13\pm0.45)\times10^{-2},
\label{pglitchtrap}
\end{equation}
\begin{equation}
\delta\Omega_{\rm s}=(6.91\pm 2.07)\times10^{-3} \mbox{\rads}.
\end{equation}

\begin{figure}
\centering
\vspace{0.1cm}
\includegraphics[width=1.0\linewidth]{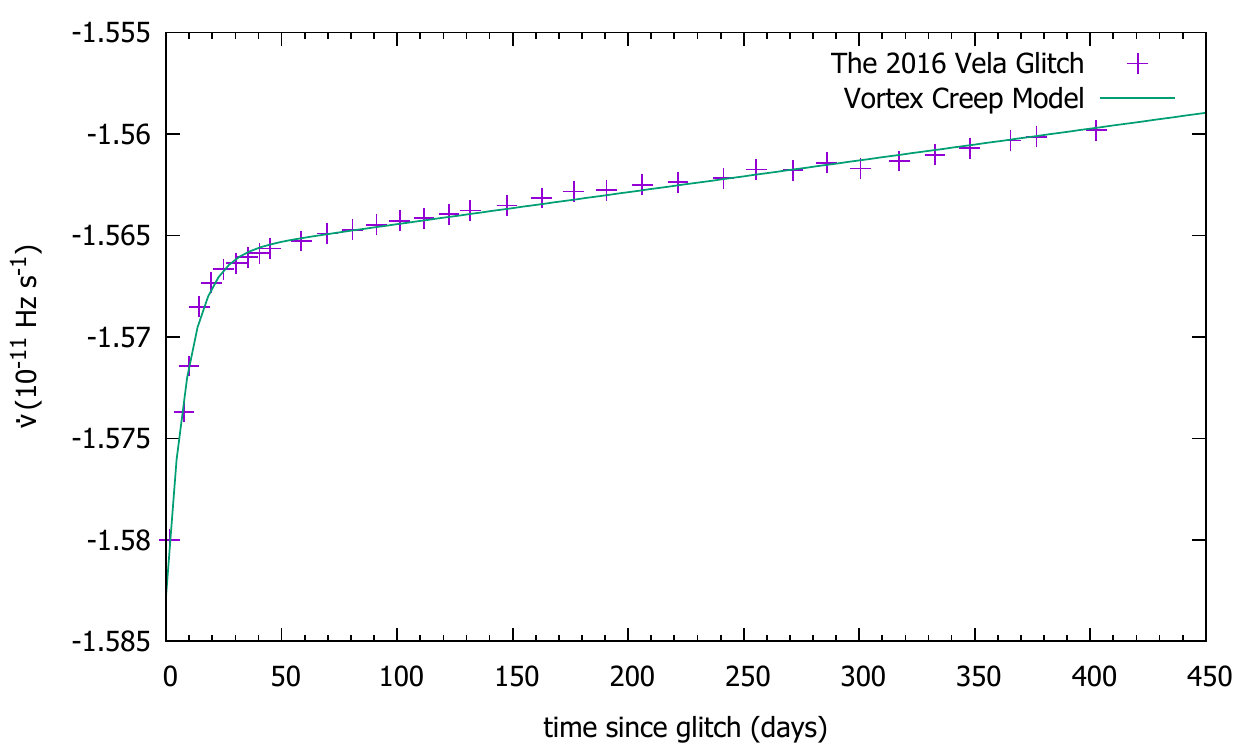}
\caption{Comparison between the 2016 Vela post-glitch data by \citet{xu19} (purple) and the vortex creep model (green).}
\label{modelfit}
\end{figure}

We use these parameter values as determined from the long term timing behaviour \citep{xu19} as input in a treatment 
of the full range of timing data \citep{palfreyman18,ashton19} starting from the 12.6 s gap containing the 
actual occurrence of the glitch. 
In the vortex creep model the post-glitch behaviour of the spin-down rate can be expressed as \citep{alpar96,erbil17a}
\begin{align}
\Delta\dot\nu(t)=&-\frac{I_{\rm tor}}{I_{\rm c}}\frac{\Delta\nu}{\tau_{\rm tor}}e^{-t/\tau_{\rm tor}}+\nonumber \\& \frac{I_{\rm A}}{I_{\rm c}}\dot\nu_{\infty}\left(1-\frac{1-(\tau_{\rm nl}/t_{0})\ln\left[1+(e^{t_{0}/\tau_{\rm nl}}-1)e^{-t/\tau_{\rm nl}}\right]}{1-e^{-t/\tau_{\rm nl}}}\right),
\label{creepmodel}
\end{align}
where non-linear creep relaxation time-scale is given by \citep{alpar84}
\begin{equation}
\tau_{\rm nl}\equiv \frac{kT}{E_{\rm p}}\frac{\omega_{\rm cr}}{\vert\dot{\Omega}\vert},
\label{taunl}
\end{equation}
and the recoupling time-scale for vortex creep against pinning to flux lines in the toroidal field region \citep{erbil14} is
\begin{align}
\tau_{\rm tor}\simeq60&\left(\frac{\vert\dot{\Omega}\vert}{10^{-10}\mbox{
\radss}}\right)^{-1}\left(\frac{T}{10^{8}\mbox{
K}}\right)\left(\frac{R}{10^{6}\mbox{ cm}}\right)^{-1}
x_{\rm p}^{1/2}\times\nonumber\\&\left(\frac{m_{\rm p}^*}{m_{\rm p}}\right)^{-1/2}\left(\frac{\rho}{10^{14}\mbox{\gcc}}\right)^{-1/2}\left(\frac{B_{\phi}}{10^{14}\mbox{ G}}\right)^{1/2}\mbox{days,}
\label{tautor}
\end{align}
with $B_{\phi}$ is the magnitude of the toroidal component of the magnetic field.
In Eq. (\ref{creepmodel}) the second term in the parenthesis reduces to $\ddot\nu t$ for $t\gtrsim \tau_{\rm nl}$. 
The exact expression in Eq.(\ref{creepmodel}) provides a better fit at intermediate 
time-scales $t \lesssim \tau_{\rm nl} \cong 45$ days. 
The vortex creep model fit to the data is shown in Figure \ref{modelfit}. Model parameters obtained from the fit are given in Table \ref{fitmodel}. 

$I_{\rm B}/I $ is interpreted as the fractional size of the vortex free region. This parameter inferred from post-glitch timing 
data agrees with our estimate above, in Eq. (\ref{newtrap}), for the new vortex trap formed just before the glitch, 
obtained from the pre-glitch slow-down episode. The total crustal superfluid moment of inertia participated in the 2016 Vela glitch 
is $I_{\rm s}=I_{\rm A}+I_{\rm B}= 1.68(4)\times10^{-2}I_{\rm c}$.

\begin{table}
\centering
\caption{Post-glitch parameter values from data fit with Eq.(\ref{modeleq}) by \citet{xu19} and inferred parameters from Eq. (\ref{creepmodel}) with the vortex creep model. The vortex creep model fit to the data from \citet{xu19} is shown in Figure \ref{modelfit}.}
\begin{tabular}{lll}
\hline\hline
Parameter & Value & Reference \\
\hline
$\Delta\nu_{\rm p} (\mbox{Hz})$ & 1.60085(9)$\times10^{-5}$ & \citet{xu19} \\
$\Delta\dot\nu_{\rm p} (\mbox{\ss})$ & -1.0(3)$\times10^{-13}$ & \citet{xu19} \\
$\Delta\ddot\nu_{\rm p} (\mbox{\sss})$ & 1.416(13)$\times10^{-21}$ & \citet{xu19} \\
$\Delta\nu_{\rm d1} (\mbox{Hz})$ & 7.7(5)$\times10^{-8}$ & \citet{xu19} \\
$ \tau_{\rm d1} (\rm days) $ & 0.96(17) & \citet{xu19} \\
$\Delta\nu_{\rm d2} (\mbox{Hz})$ & 6.05(7)$\times10^{-8}$ & \citet{xu19} \\
$ \tau_{\rm d2} (\rm days) $ & 6.0(5) & \citet{xu19} \\
$ t_{\rm g} (\rm days) $ & 782 & \citet{gancio20} \\
$ I_{\rm A}/I_{\rm c} $ & 3.48(21)$\times10^{-3}$ & this work \\
$ I_{\rm B}/I_{\rm c} $ & 1.33(5)$\times10^{-2}$ & this work \\
$ I_{\rm tor}/I_{\rm c} $ & 1.82(49)$\times10^{-2}$ & this work \\
$ \tau_{\rm nl} (\rm days) $ & 45(10) & this work \\
$ \tau_{\rm tor} (\rm days) $ & 9(1) & this work \\
$ t_{0} (\rm days) $ & 781(13) & this work \\
\hline
\label{fitmodel}
\end{tabular}
\end{table} 

\section{Discussion and Conclusions} \label{sec:dandc}

The vortex unpinning and creep model for pulsar glitches involves angular momentum exchange between three components 
of the neutron star. The glitch itself is due to sudden unpinning and outward motion of vortices in the crustal superfluid 
as a result of an instability which is still not understood in detail. This event first transfers angular momentum to the crust normal 
matter, a solid lattice of nuclei which can pin and interact with vortices. The angular momentum is then shared  
with the core superfluid via electrons. After these two initial short time-scale angular momentum exchanges the core superfluid 
is effectively part of the crust and normal matter system to which it is tightly coupled. 
Finally the vortex creep process, offset by the glitch induced changes in the rotation rates of the crustal superfluid and 
the effective crust, relaxes back towards steady-state conditions on post-glitch and inter-glitch 
timescales of days to a few years. 

In all glitches observed from the Vela and other pulsars prior to the 
Vela pulsar's 2016 glitch, the initial angular momentum transfer from the crustal superfluid to the crust normal matter, 
and then to the core superfluid were not resolved by the observations, which were modelled in terms of 
only two components; the crustal superfluid and the effective crust including the core superfluid.
Observation of the 2016 Vela glitch by \citet{palfreyman18} and its subsequent reanalysis by \citet{ashton19} 
resolved the early timing signatures of the glitch for the first time. 

Just prior to the 2016 glitch the rotation rate of the Vela pulsar was found to decrease. We have evaluated 
this observation as an indication of formation of new vortex traps by a crustquake which thereafter triggers the vortex 
unpinning avalanche that constitutes the glitch. Around the glitch ephemeral changes in the 
pulsar's electromagnetic signature were detected, which we interpret as the aftermath of the crust breaking event 
on the magnetospheric conditions. 

The 2016 Vela glitch has revealed the best 
constraint on the spin-up time-scale of the observed crust, placing 
an upper limit of 12.6 seconds for the glitch rise time. We have interpreted this as an upper limit on the time-scale 
of angular momentum transfer from the unpinned vortices to the nuclei forming the crust solid. 
After the peak glitch spin-up the crustal rotation rate 
promptly relaxed within a minute. This is interpreted in our scenario as the gradual coupling of the core superfluid to the normal 
matter crust via the electrons' scattering off magnetised vortex lines. 

These observations yield  important information 
on the neutron star internal structure and dynamics. With the 2016 Vela pulsar glitch we are able for the first time 
to discuss the sequence of angular momentum transfer 
between the three components, the crustal superfluid, the crust normal matter (and electrons throughout the star), 
and the core superfluid.  We have taken the sequence of events in chronological order, 
starting with (i) the formation of a vortex trap that triggered the glitch, 
proceeding through (ii) short-lived magnetospheric changes associated with crust breaking at the glitch, 
which (iii) transfers angular momentum from crustal superfluid 
to the solid crust via the interactions of unpinned vortices with the nuclei forming the crust lattice. This is followed by (iv) 
the gradual coupling of the core superfluid to the crust and finally by (v) the longer term relaxation of the crustal superfluid, 
which is the only process addressed by modelling the post-glitch and interglitch timing data for previous glitches. 
We make the following inferences for this sequence of developments:

(i) There are strong indications that formation of new vortex traps, 
coupled to breaking of the solid crust, may be triggering the vortex unpinning glitch. Glitch associated changes in the 
electromagnetic signature of PSR J1119-6127 \citep{akbal15} and the early  timing data from the largest Crab pulsar glitch 
\citep{erbil19} were interpreted in terms of such crust breaking. 
There were no indications of a trigger event in glitch associated timing data of the Vela pulsar prior to the 2016 glitch. 
But the precursor slow down event extending over 100 seconds just prior to this glitch \citep{ashton19} 
is below the resolution of timing observations for the previous glitches. 
We have evaluated this pre-glitch decrease in the crustal rotation rate in terms of formation of a new vortex trap 
which initiated the glitch. In a crustal superfluid region vortex creep is continuously transferring angular momentum to the crust. 
If a vortex trap forms as a result of some superfluid hydrodynamical instability and/or an agent like crust breaking 
in a quake, then the angular momentum transfer from 
this new trap ceases irreversibly so that a decrease in the crustal rotation rate follows. The time-scale $\tau_{\rm tr} =[ (\ell_{\rm v}/v_{0})\exp(E_{\rm p}/{kT})]$ of 
the decline is the time-scale of re-establishing the pinned vortex distribution, estimated as the vortex transit time 
over the inter-vortex spacing $\ell_{\rm v}$ at the microscopic vortex speed $v_{0}$ taking into account the pinning and unpinning process. 
In the inner crust, for $\rho\leq10^{14}\mbox{\gcc}$ this vortex transit time agrees with  
$\simeq 100$ seconds, the observed duration of the slow-down event before the 2016 glitch. 
The formation of new vortex traps also triggers the collective unpinning event that is observed as the glitch. This means 
that the fluctuations in the vortex number density and local superfluid velocity which arise during the vortex trap formation process 
raise the angular velocity lag between the crustal superfluid and the crust normal matter from its steady-state creep value 
to the critical value for vortex unpinning and thus initiate the glitch. Estimating the change in the local lag 
and in the superfluid rotation rate, together with the observed pre-glitch slow-down of the crust one obtains 
the fractional moment of inertia of the newly formed trap regions to be  $I_{\rm trap}/I=8.58\times10^{-3}$.
Since the vortex unpinning avalanche starts at moderate depths corresponding to our density 
estimate of $\rho\leq10^{14}\mbox{\gcc}$ and the outward motion of unpinned vortices does not 
cover the entire crustal superfluid, the glitch magnitude is not expected to attain a large value. 
This is indeed the case as the 2016 glitch is amongst the moderate size events observed for the Vela \citep{xu19}. 

(ii) This is the first Vela glitch resolved to display changes in electromagnetic signature of the pulsar. 
These changes were recently addressed by \citet{bransgrove20} in terms of a wave transmission model 
for seismic activity deep inside the crust which 
released energy to high frequency magnetospheric modes and induced temporary electron/positron discharge in the magnetosphere. 
There are very few previous examples of glitch associated changes in the pulsar signature. In PSR J1119-6127 
emergent additional pulse components extending to about three months were observed following the 2007 glitch. 
\citet{akbal15} interpreted these observations in terms of a quake involving crust plates extending to the surface which 
bring about low frequency plastic motion of the magnetic field lines on the scale size of the broken plate, $D\sim10-100$ m, 
and their subsequent relaxation to a new configuration on a three month time-scale. For the 2016 Vela glitch, a quake 
occurring deep inside the crust is invoked to induce high 
frequency oscillations leading to the observed changes of duration $\simeq$4.4 s  in the magnetosphere, 
while most of the released quake energy is drained to the core \citep{bransgrove20}. 
The underlying physical reason for the qualitative differences in glitch associated pulsar behaviour between the 2016 
Vela glitch and the 2007 glitch of PSR J1119-6127 is the location of the quake. A quake in the inner crust is 
indicated for the 2016 glitch so that its effects reach the surface and magnetospheric field lines anchored in the surface via 
high frequency elastic waves as envisaged by \citet{bransgrove20}. This is consistent with our inference of an  inner crust 
location, at a density $\rho\leq 10^{14}\mbox{\gcc}$  for the formation of vortex traps via a quake. 

(iii) The observed 12.6 seconds upper limit for the glitch rise time has implications for the efficiency of angular momentum 
exchange mechanism coupling the crustal superfluid to the normal matter in the crust. 
As the unpinned vortices move through the crust, they interact with crustal nuclei and phonons via kelvons which are 
excitations of the vortices. We have obtained coupling times   
 $\tau_{\rm mf-kelvon}\sim0.1-1$ s by applying the results of \citet{graber18} in Eqs. (\ref{kt}) and (\ref{dragk}) 
 for a range of values of the microscopic vortex velocity around nuclei calculated by \citet{erbil16}. 
 The 12.6 s  upper limit for the glitch rise time is not likely to be improved substantially in future glitches 
 since timing analysis is fundamentally limited by the requirement of pulse templates to be 
 constructed from a train of many individual pulses. The coupling time $\tau_{\rm mf-kelvon}\sim0.1-1$ s is therefore 
 unlikely to be resolved in glitches electromagnetically. An interesting speculation is that $\tau_{\rm mf-kelvon}\sim0.1-1$ s might 
 be detected with future gravitational wave detectors if it sets the scale for transient gravitational wave emission associated with glitches.

(iv) The peak glitch spin up observed at 12.6 s after the last pre-glitch data continues to relax as the crustal superfluid-normal matter system transfers angular momentum back to the core superfluid. The coupling is mediated by magnetized vortex line-electron scattering \citep{ALS84}.  
The coupling time $\tau_{\rm core}$ as a function of density is shown in Figure \ref{ALScoupling}. We find that the core superfluid regions with 
densities $\gtrsim7.4\times 10^{14}$\gcc~with coupling times $\tau_{\rm core}$ less than $12.6$ seconds  
were already coupled to the normal matter crust when the glitch was resolved. The prompt 
relaxation immediately following the peak spin-up, which has an effective relaxation time 
$\tau_{\rm d+}=53.96^{+24.02}_{-14.82}$ s, and associated fractional moment of inertia $I'_{\rm core}\cong(0.38-0.66)I_{\rm c}$ are shown in Figure \ref{ALScoupling}. This response corresponds to the gradual coupling of the core regions at densities 
$2.7\lesssim \rho(10^{14}\mbox{\gcc})\lesssim3.8$. The coupling time 
$\tau_{\rm core}$ given by Eq.(\ref{ALStime}) corresponds to the observed $\tau_{\rm d+}=53.96^{+24.02}_{-14.82}$ s for effective proton masses in the range $m^{*}_{\rm p}/m_{\rm p} \cong (0.56-0.63)$ supporting theoretical estimates in this density range for various EoSs \citep{chamel08}.

(v) The long term timing behaviour after the core superfluid and the crustal normal matter are fully coupled is the 
part of the post-glitch process that was like observed from the previous glitches. The characteristic post-glitch and inter-glitch 
behaviour following all Vela pulsar glitches is ascribed to the recovery of the vortex creep process. 
\citet{chau93} fitted the first 9 glitches of the Vela pulsar with a function including two exponentially decaying terms, 
with time constants of 3.2 and 33 days, and a long term decrease of the spin-down rate with constant second derivative 
$\Delta\ddot{\nu}_{\rm p}$, in the form $\Delta\dot\nu_{\rm p}(1-t/t_{0})$. The exponential decays are interpreted as the 
linear response of certain creep regions, whose fractional moments of inertia are the fractional amplitudes in spin-down rate 
of the corresponding exponentially decaying terms. The constant second derivative term is the highly non-linear response of creep 
from crustal superfluid regions whose moment of inertia is extracted from the 
fits as $I_{\rm A}/I = \Delta\dot\nu_{\rm p}/\dot\nu$. This form is an approximation which is not valid on the short time-scales 
within a few days after the glitch.
In the 9 early Vela glitches the uncertainties in the date of the glitch were a few days or longer and immediate post-glitch 
data were sparse or lacking. Applying the same terms employed by \citet{chau93} 
to the 2016 Vela glitch does not fit the earliest data points well, and also leads to an increase in moments of 
inertia of the corresponding creep components that is hard to accommodate with crustal superfluid alone. 
We have found that the exact non-linear creep response given in Eq. (\ref{creepmodel}) 
plus a single exponentially decaying component with time constant $\tau_{\rm tor}=9$ days satisfactorily 
describes the 2016 post-glitch data. We associate this term with the response of vortex creep 
against toroidal flux tubes of the proton superconductor in the outer core \citep{erbil17a}. 
The amplitude of this term constrains the extent of the toroidal field region 
to a fractional moment of inertia of $ I_{\rm tor}/I=1.82\times10^{-2}$.
The 9 day relaxation time gives information about the interaction between the superfluid vortex lines and 
the magnetic flux tubes. The dominant response of the core superfluid-proton superconductor system to the long term post-glitch recovery comes from the regions with $\rho\lesssim8\times10^{14}$\gcc. We estimate the critical lag for unpinning of vortex lines from magnetic flux tubes as $\omega_{\rm cr, v-\Phi}\approx 0.08$\rads~and vortex line-flux tube pinning energy per junction as  $E_{\rm p, v-\Phi}\approx 2$ MeV \citep{erbil16}. The range of densities inferred for the outer core toroidal field region 
is consistent with the range $\rho\gtrsim3\times 10^{15}$\gcc~from Eq.(\ref{tautor}) inferred for the superfuid (inner) core wherein there are no toroidal flux tubes, pinning and creep. 

The non-linear creep response term gives the crustal superfluid recoupling time-scale $\tau_{\rm nl}=45$ days. For $E_{\rm p}=0.17$ MeV and $kT=6.5$ keV this non-linear creep time-scale implies $\omega_{\rm cr}\cong 0.01$\rads~by Eq.(\ref{taunl}), characterizing conditions for superweak pinning in the inner crust \citep{alpar84,alpar89}. 

The non-linear creep response term also yields the prediction $t_{0}=781$ days for the time to the next glitch. This estimate matches the observed 
interval $t_{\rm g}=782$ days \citep{sarkissian19,gancio20} from the 2016 to the 2019 glitch quite well. \citet{chau93} and \citet{akbal17} compared the
non-linear creep model estimates of the times to the next glitch obtained from the parameters of the previous glitch with the observed inter-glitch intervals, for all 16 large Vela glitches observed up to 2016. For that sample, the deviation of the creep model prediction $t_{\rm 0i}$ for the time to the next glitch from the observed $t_{\rm gi}$ inter-glitch time of the $i^{\rm th}$ glitch $(t_{\rm 0i}-t_{\rm gi})/t_{\rm gi}$ is at most 35 percent and the agreement between the model prediction and the observations improves when a glitch associated persistent shift in the spin-down rate is included \citep{akbal17}. Our estimate $t_{0}\cong t_{\rm g}$ for the 2016 glitch within 1 day has the closest agreement among the entire sample of 17 inter-glitch intervals up to the 2019 glitch. Thus, our estimate strengthens the applicability of the vortex creep model for predicting the date of future glitches from the parameters of the previous one within reasonable errors. 

The post-glitch $\dot\nu(t)$ data of the 2019 Vela glitch have not been published yet \citep{lower20,gancio20}. Once these data become available we will be able to determine the properties of the superfluid layers involved in the 2019 glitch and make a prediction for the time to the next glitch by using the glitch magnitude and the glitch associated change
in the spin-down rate remaining after the immediate post-glitch exponential relaxations are over.

Nevertheless, we can draw a general picture on the basis of basic model parameters \citep{pines80,alpar81}: There exists an anti-correlation between $I_{\rm A}$ and $\delta\Omega_{\rm s}$ so that $\Delta\Omega_{\rm c}$ remains approximately the same within a factor of few in all large Vela glitches. This fact is understandable, since if in a glitch the resulting lag $\omega_{\infty}-\delta\Omega_{\rm s}$ is large (small) [that is change in superfluid velocity $\delta\Omega_{\rm s}$ is small (large)] unpinned vortices are scattered much (less) and vortex lines travel a larger (shorter) distance so that the affected region $I_{\rm A}+I_{\rm B}$ ends up with larger (smaller) for that glitch. And if  $\omega_{\infty}-\delta\Omega_{\rm s}$ is large (small) in a glitch, then creep will be more (less) effective with vortex accumulation proceeds slower (faster) so that the next glitch will be larger (smaller). This is consistent with the observations \citep{lower20} that somewhat smaller ($\Delta\nu/\nu\cong 1.44\times10^{-6}$) 2016 Vela glitch is followed by a larger ($\Delta\nu/\nu\cong 2.5\times10^{-6}$) 2019 glitch. Also on this basis we expect that the next large Vela glitch will be smaller than the 2019 glitch provided that $t_{\rm g,2019}\gtrsim t_{\rm g,2016}$.

The total crustal superfluid moment of inertia that participated in the 2016 Vela glitch is $I_{\rm s}/I=1.52(3)\times10^{-2}$,
below the constraint brought by crustal entrainment effect if the crust is crystalline even for quite large mass entrainment enhancement factor of neutrons $<m^*_{\rm n}/m_{\rm n}>=5.1$ \citep{delsate16}, for a 1.4M$_{\odot}$ neutron star with a thick crust \citep{basu18}. [If the neutron star 
crust is actually disordered the entrainment effect for the crust superfluid does not bring a significant constraint on the crustal superfluid moment of inertia \citep{sauls20}.] 

The model presented here makes the following predictions for the Vela glitches: 
\begin{enumerate}
\item If a quake is responsible for the formation of vortex traps that initiates the large scale vortex unpinning avalanche and triggers a glitch, a slow-down episode prior to the glitch associated with the newly formed vortex traps will be visible only if glitch starts at not so deep places of the crust, i.e. for $I_{\rm A}/I\lesssim7\times10^{-3}$ (Because smaller glitches originate at lower density layers of the inner crust where the trap formation time-scales are longer).
\item If the glitch is triggered by crust breaking, this will induce short-lived transient magnetospheric changes like pulse shape and polarization level variations. Deeper the place of the quake (the larger the size of the glitch), shorter the duration of such events. 
\item A glitch will not be caught really in the act because the initial angular momentum exchange between vortex lines and lattice nuclei takes place on extremely short time-scales of $\tau_{\rm rise}\lesssim2$ s.
\item If resolved, a Vela glitch should start with an initial peak in the crustal rotational velocity followed by a quick relaxation, reflecting the gradual coupling of the core superfluid to the crustal superfluid-crust normal matter system.
\item The long term post-glitch spin-down rate evolution after the early exponential recoveries are over will be characterized by a large, constant, positive second derivative of the rotation rate. These long-term timing data provide a rough estimate for the time to the next glitch. 
\end{enumerate}
  
\section*{Acknowledgements}
\addcontentsline{toc}{section}{Acknowledgements}
This work is supported by the Scientific and Technological Research Council of Turkey
(T\"{U}B\.{I}TAK) under the grant 117F330. M.A.A. is a member of the Science Academy
(Bilim Akademisi), Turkey. We thank the referee for particularly useful comments which lead to clarification of some points.


\bsp	
\label{lastpage}

\begin{thebibliography}{60}

\bibitem[Akbal et al.(2017)]{akbal17}
Akbal O., Alpar M.~A., Buchner S., \& Pines D.,\ 2017, \mnras, 469, 4183 
\bibitem[Akbal et al.(2015)]{akbal15} Akbal O., G\"{u}gercino\u{g}lu 
E., \c{S}a\c{s}maz Mu\c{s} S., \& Alpar M.~A.,\ 2015, \mnras, 449, 933
\bibitem[Akmal et al.(1998)]{akmal98} Akmal, A., Pandharipande, 
V.~R., \& Ravenhall, D.~G.\ 1998, \prc, 58, 1804 
\bibitem[Alpar(1977)]{alpar77}
Alpar M.~A., \ 1977, \apj, 213, 527
\bibitem[Alpar et al.(1981)]{alpar81}
Alpar M. A., Anderson P. W., Pines D., \& Shaham J., \ 1981, \apjl, 249, L29
\bibitem[Alpar et al.(1984a)]{alpar84}
Alpar M. A., Anderson P. W., Pines D., \& Shaham J., \ 1984a, \apj, 276, 325
\bibitem[Alpar \& Baykal(2006)]{alpar06}
Alpar M~.A., \& Baykal A.,\ 2006, \mnras, 372, 489
\bibitem[Alpar et al.(1996)]{alpar96}
Alpar M. A., Chau H. F., Cheng K. S., \& Pines D., 1996, ApJ, 459, 706
\bibitem[Alpar et al.(1989)]{alpar89}
Alpar M. A., Cheng K. S., \& Pines D., \ 1989, \apj, 346, 823
\bibitem[Alpar et al.(1984b)]{ALS84}
Alpar M. A., Langer, S.A., \& Sauls, J.A., \ 1984b, \apj, 282, 533
\bibitem[Anderson \& Itoh(1975)]{anderson75}
Anderson P.~W., \& Itoh N.,\ 1975, \nat, 256, 25
\bibitem[Antonopoulou et al.(2015)]{antonopoulou15}
Antonopoulou D., Weltevrede P., Espinoza C.~M., Watts A.~L., Johnston S., Shannon R.~M., \& Kerr M., \ 2015, \mnras, 447, 3924
\bibitem[Ashton et al.(2019)]{ashton19}
Ashton G., Lasky P.~D., Graber V., \& Palfreyman J., \ 2019, Nature Astronomy, 3, 1143
\bibitem[Basu et al.(2018)]{basu18}
Basu A., Char P., Nandi R., et al.\ 2018, \apj, 866, 94
\bibitem[Basu et al.(2020)]{basu20}
Basu A., Joshi B.~C., Krishnakumar M.~A., et al.\ 2020, \mnras, 491, 3182
\bibitem[Bransgrove et al.(2020)]{bransgrove20}
Bransgrove A., Beloborodov A.~M., \& Levin Y.\ 2020, arXiv:2001.08658
\bibitem[Buchner \& Flanagan(2008)]{buchner08}
Buchner S., \& Flanagan C., 2008, in Bassa C., Wang Z., Cumming A., Kaspi V.
M., eds., AIP Conf. Proc. Ser. Vol. 983, 40 Years of Pulsars:Millisecond
Pulsars, Magnetars and More. Am. Inst. Phys., New York, p. 145
\bibitem[Celora et al.(2020)]{celora20}
Celora T., Khomenko V., Antonelli M., \&  Haskell B., \ 2020, arXiv:2002.04310
\bibitem[Chamel(2008)]{chamel08} Chamel N.,\ 2008, \mnras, 388, 737
\bibitem[Chau et al.(1993)]{chau93}
Chau H.~F., McCulloch P.~M., Nandkumar R., \& Pines D., 1993, \apjl, 413, L113
\bibitem[Cheng et al.(1988)]{cheng88}
Cheng K.~S., Alpar M.~A., Pines D. \& Shaham J., \ 1988, \apj, 330, 835
\bibitem[Cordes et al.(1988)]{cordes88}
Cordes J.~M., Downs G.~S. \& Krause-Polstorff J., \ 1988, \apj, 330, 847
\bibitem[Dang et al.(2020)]{dang20}
Dang S.~J., Yuan J.~P., Manchester R.~N., et al.\ 2020, \apj, 896, 140
\bibitem[Delsate et al.(2016)]{delsate16}
Delsate T., Chamel N., G{\"u}rlebeck N., Fantina A.~F.,  Pearson J.~M., \& Ducoin C.,\ 2016, \prd, 94, 023008
\bibitem[Dodson et al.(2007)]{dodson07}
Dodson R.~G., Lewis D.~R., \& McCulloch P.~M., \ 2007, \apss, 308, 585
\bibitem[Dodson et al.(2002)]{dodson02}
Dodson R.~G., McCulloch P.~M., \&  Lewis D.~R., \ 2002, \apjl, 564, L85
\bibitem[Epstein \& Baym(1992)]{epstein92}
Epstein R.~I., \& Baym G.,\ 1992, \apj, 387, 276
\bibitem[Gancio et al.(2020)]{gancio20}
Gancio G., Lousto C.~O., \& Combi L., et al., \ 2020, \aap, 633, A84
\bibitem[Gavassino et al.(2020)]{gavassino20}
Gavassino L., Antonelli M., Pizzochero P.~M.,  \& Haskell B. \ 2020, \mnras, 494, 3562
\bibitem[Graber et al.(2018)]{graber18}
Graber V., Cumming A., \& Andersson N., \ 2018, \apj, 865, 23
\bibitem[G{\"u}gercino{\u g}lu(2017)]{erbil17a}
G{\"u}gercino{\u g}lu E., \ 2017, \mnras, 469, 2313
\bibitem[G{\"u}gercino{\u g}lu \& Alpar(2014)]{erbil14}
G{\"u}gercino{\u g}lu E., \& Alpar M.~A.,\ 2014, \apjl, 788, L11
\bibitem[G{\"u}gercino{\u g}lu \& Alpar(2016)]{erbil16}
G{\"u}gercino{\u g}lu E., \& Alpar M.~A.,\ 2016, \mnras, 462, 1453
\bibitem[G{\"u}gercino{\u g}lu \& Alpar(2019)]{erbil19}
G{\"u}gercino{\u g}lu E., \& Alpar M.~A.,\ 2019, \mnras, 488, 2275
\bibitem[Johnston \& Galloway(1999)]{johnston99} Johnston S., \& Galloway D.,\ 1999, \mnras, 306, L50
\bibitem[Jones(1992)]{jones92} Jones, P.~B.\ 1992, \mnras, 257, 501
\bibitem[Kaspi \& Beloborodov(2017)]{kaspi17} Kaspi, V.~M., \& Beloborodov, A.,\ 2017, \araa, 55, 261
\bibitem[Keitel et al.(2019)]{keitel19}
Keitel D., Woan G., Pitkin M., et al.\ 2019, \prd, 100, 064058
\bibitem[Lower et al.(2020)]{lower20}
Lower M., Bailes M., Shannon R.~M., et al.\ 2020, \mnras, 494, 228
\bibitem[L\"onnborn et al.(2019)]{melatos19}
L\"onnborn J.~R., Melatos A., \& Haskell B.\ 2019, \mnras, 487, 702
\bibitem[Lyne et al.(1993)]{lyne93}
Lyne, A.~G. Pritchard, R.~S., \& Graham Smith, F., \ 1993, \mnras, 265, 1003
\bibitem[Melatos et al.(2015)]{melatos15}
Melatos A., Douglass J.~A., \& Simula T.~P.\ 2015, \apj, 807, 132
\bibitem[Palfreyman et al.(2018)]{palfreyman18}
Palfreyman J., Dickey J.~M., Hotan A., Ellingsen S., \& van Straten W., \ 2018, \nat, 556, 219
\bibitem[Pines et al.(1980)]{pines80}
Pines D., Shaham J., Alpar M.~A., \& Anderson P.~W., \ 1980, \ptp, 69, 376
\bibitem[Pizzochero et al.(2020)]{pizzochero19}
Pizzochero P., Montoli A., \& Antonelli M., \ 2020,  \aap, 636, A101
\bibitem[Sarkissian et al.(2019)]{sarkissian19}
Sarkissian J., Hobbs G., Reynolds J., Palfreyman J., \&  Olney S.,  \ 2019,  The Astronomer's Telegram, 12466, 1
\bibitem[Sauls et al.(2020)]{sauls20}
Sauls J.~A., Chamel N., \& Alpar M.~A., \ 2020, arXiv: 2001.09959
\bibitem[Seveso et al.(2016)]{seveso16}
Seveso S., Pizzochero P.~M., Grill F., \& Haskell B.\ 2016, \mnras, 455, 3952 
\bibitem[Shannon et al.(2016)]{shannon16}
Shannon R.~M., Lentati L.~T., \& Kerr, M., et al., \ 2016, \mnras, 459, 3104 
\bibitem[Sidery \& Alpar(2009)]{sidery09} Sidery T., \& Alpar M.~A.,\ 2009, \mnras, 400, 1859
\bibitem[Sourie \& Chamel(2020)]{sourie20} Sourie A., \& Chamel N.,\ 2020, \mnras, 493, L98
\bibitem[Sourie et al.(2017)]{sourie17} Sourie A., Chamel N., Novak J., \& Oertel M.\ 2017, \mnras, 464, 4641
\bibitem[Vigano et al.(2013)]{vigano13}
Vigano D., Rea N., Pons J.~A., et al., \ 2013, \mnras, 434, 123
\bibitem[Warszawski \& Melatos(2011)]{warszawski11}
Warszawski L., \& Melatos A.\ 2011, \mnras, 415, 1611 
\bibitem[Weltevrede et al.(2011)]{weltevrede11}
Weltevrede P., Johnston S., \& Espinoza C.~M., \ 2011, \mnras, 411, 1917
\bibitem[Xu et al.(2019)]{xu19}
Xu Y.~H., Yuan J.~P., Lee K.~J., et al., \ 2019, \apss, 364, 11
\bibitem[Yu et al.(2013)]{yu13}
Yu M., Manchester R.~N., Hobbs G., et al.\ 2013, \mnras, 429, 688
\bibitem[Yuan et al.(2019)]{yuan19}
Yuan J.~P., Kou F.~F., \& Wang N., \ 2019, in Ang Li, Bao-An Li and Furong Xu, eds., Xiamen-CUSTIPEN Workshop on the Equation of State of Dense Neutron-Rich Matter in the Era of Gravitational Wave Astronomy, AIP Conf. Proc. Ser. Vol. 2127, 020004
\end{thebibliography}
\end{document}